\documentclass[twocolumn]{revtex4}

\usepackage{graphicx}
\usepackage{dcolumn}
\usepackage{amsmath}
\begin{document}

%\preprint{}

\title{A group theoretic approach to shear-free radiating stars}



\author{G. Abebe}\email[]{gezahegn@aims.ac.za}
\author{S. D. Maharaj}\email[]{maharaj@ukzn.ac.za}
\author{K. S Govinder}\email[]{govinder@ukzn.ac.za}

\affiliation{Astrophysics and Cosmology Research Unit, School of Mathematics, Statistics and Computer Science, 
University of KwaZulu--Natal, Private Bag X54001, Durban 4000, South Africa}

%\thanks{}
%\date{\today}

\begin{abstract}
A systematic analysis of the junction condition,  relating the radial pressure with the heat flow in a shear-free
relativistic  radiating star, is undertaken.  This is a highly
nonlinear partial differential equation in general.
We obtain the Lie point symmetries that leave the boundary condition invariant. Using a linear combination of the symmetries, we transform the junction condition into ordinary differential equations. We present several new exact solutions to the junction condition. In each case we can identify the exact  solution with a Lie point generator. Some of the solutions  obtained  satisfy the linear barotropic equation of state. As a special case we regain conformally flat models which were found previously. 
Our analysis highlights the interplay between Lie algebras, nonlinear differential equations and application to relativistic astrophysics.
\end{abstract}

%\pacs{04.20.Jb, 04.20.Nr, 04.70.Bw}

\maketitle


\section{Introduction}
 Radiating stars in a general relativistic context have been widely studied  because of their importance  in astrophysics. The detailed physics of such models have been investigated by Herrera \emph{et al.} \cite{hh1} with dissipation, by Di Prisco \emph{et al.} \cite{hh2} with charge, and Herrera \emph{et al.} \cite{hh3} with viscous dissipation in  casual thermodynamics in the streaming and diffusion approximations.  Particular models have been recently studied by Chan \emph{et al.} \cite{hh4} for an imperfect nonadiabatic distribution,  by Pinheiro and Chan \cite{hh5} with charge,  and Pinheiro and Chan \cite{hh6} for a collapsing  body with an initial inhomogeneous  density. Herrera and Santos \cite{hh7} introduced the concept of Euclidean stars  in general relativity by requiring that the  areal radius and proper radius are equal. Exact  models  satisfying this condition were generated by Govender \emph{et al.} \cite{hh8} and Govinder and Govender \cite{hh9}. Abebe \emph{et al.} \cite{b12} showed  that generalized Euclidean stars may be generated  with an equation of state.  An interesting feature of their approach is the application of Lie symmetries to produce models which are expanding, accelerating and shearing. This indicates that the Lie symmetry approach is helpful in solving the boundary junction condition, and highlights its role in astrophysical applications. We point out that in recent times  Lie symmetry generators have been particularly useful in generating  models in spherically symmetric gravitational fields. Msomi \emph{et al.}  \cite{b4,b5} have considered applications with heat flow and generated new classes of exact solutions.  Kweyama \emph{et al.}  \cite{b7, b77} analysed Noether and Lie symmetries for charged relativistic fluids and were able to solve the field equations exactly.

 The shear-free assumption in  stellar models is often  made in the study of relativistic radiating stars. Kolassis \emph{et al.} \cite{a} obtained  the first exact solution with dissipation effects, satisfying the boundary conditions, for a shear-free radiating star in geodesic motion.  
Thirukkanesh and Maharaj \cite{e} generated exact solutions for  the geodesic model by transforming the junction condition into Bernoulli, Riccati, and confluent hypergeometric equations. This model was  extended by  Govender and Thirukkanesh \cite{f} to include a nonzero cosmological constant.  Note that in  these treatments  fluid particles travel along geodesics and the  Friedmann dust solution in the absence of heat flow was regained in all cases. An important point here is  that these treatments were performed under the condition of pressure isotropy and anisotropic pressures are absent. 
Ivanov \cite{ag} considered a general  shear-free perfect fluid  with heat flow  that contains conformal flatness and geodesic models as a special case. A conformally flat relativistic model with dissipation and inhomogeneity  was first proposed Herrera \emph{et al.} \cite{ai}. This was integrated  by Maharaj and Govender \cite{aj} and Herrera \emph{et al.} \cite{ak} to obtained   classes of exact solutions. Another  class  of conformally flat solutions was generated by Misthry \emph{et al.} \cite{al} by transforming the junction condition equation to an Abel equation. A further conformal flat radiating model, generated by a Lie symmetry was obtained by   Abebe \emph{et al.} \cite{b10}. Other models of shear-free radiating stars have been constructed by Tewari \cite{tewari1, tewari2}, Pant \emph{et al.} \cite{pant} and Pant and Tewari \cite{panttewari} which describe massive radiating fluid spheres and generated horizon-free collapse.

The shear-free assumption is utilised often in the analysis of self-gravitating spheres and when  investigating gravitational collapse.
This is largely due to the resulting simplification of the field equations. In addition, shear-free  and the homogeneous expansion rate conditions
are equivalent to the homology conditions in the Newtonian limit. Therefore the shear-free condition is well justified.
However we need to make the observation that the shear-free condition may be unstable in the presence
of anisotropic pressures and dissipative fluxes. The conditions under which an initially shear-free fluid continues to remain shear-free in its subsequent
evolution has been studied by Herrera \emph{et al.} \cite{add}. In that analysis it was demonstrated that pressure anisotropy and dissipation affect
the propagation of time and the gravitating relativistic fluid can become unstable.

 We analyse a shear-free radiating model, without assuming conformal flatness and geodesic motion of fluid particles, in the presence of pressure anisotropy.  We consider the Einstein field equations and boundary conditions in the context of Lie  symmetries with the objective of producing  new models of relativistic radiating objects.  This is the main object of this paper.  The case of shearing stellar models with Lie symmetries  has been considered in some earlier treatments. Those treatments do not allow for vanishing shear and our results cannot be regained from such investigations. The absence of shear produces fundamentally  a different set of field equations to be integrated. The Lie generators are generically different when  the shear vanishes. In Sect. \ref{335} we briefly discuss the  shear-free radiating star and  present the junction conditions.  We show that the master equation is a nonlinear partial differential equation  in the metric functions. In Sect. \ref{si3} we generate Lie symmetries for the master equation. Using the Lie point symmetries approach  we transform  the boundary condition to ordinary differential equations. By analyzing the resulting  ordinary differential equations and transforming in the original variables  we present new exact solutions in  Sect. \ref{sec4},  \ref{sec5} and \ref{sec6}.  We link our solutions to specific Lie symmetry generators. In   Sect. \ref{s5} we show that some of our solutions satisfy a linear barotropic equation of state.   We make concluding remarks, discuss  the  causal heat transport equation and summarize our solutions in tabular form in  Sect. \ref{s7}.

\section{The model\label{335}}
 We consider the particular case of spherically symmetric, shear-free radiating stellar models.  The  line element for the interior for the spacetimes is given by
 \begin{equation}\label{9}
 ds^2=-A^2dt^2+B^2\left[  dr^2+r^2\left( d\theta^2+\sin ^2\theta d\phi^2 \right) \right], 
 \end{equation}
 where  $A$ and $B$ are metric functions of  $t$ and  $r$.  The acceleration and and expansion are nonzero but the fluid is shear-free.

The energy momentum tensor has the form 
 \begin{eqnarray}
 T_{ab}&=& \left(\mu+ p_{\perp} \right) u_au_b+ p_{\perp}g_{ab}+(p_{\parallel}- p_{\perp})\chi_ {a}\chi_b \nonumber\\
  &&+q_au_b+q_bu_a, \label{73aa}
 \end{eqnarray}
with heat flux and anisotropic stress. The fluid four-velocity  $u^a=\frac{1}{A}\delta^a_0$ is comoving, $\chi^a$ is an unit four-vector along the radial direction $(u_a\chi^a=0)$, and 
 the heat flow  vector  $q^a=\left( 0,q,0,0\right)$  is radially directed $(u_aq^a=0)$.
The Einstein field equations  for the heat conducting  spherically symmetric anisotropic  fluid \eqref{73aa} become
\begin{subequations}\label{14}
 \begin{eqnarray}
 \mu&=&\frac{3}{A^2}\frac{B^2_t}{B^2}-\frac{1}{B^2}\left(2\frac{B_{rr}}{B}-\frac{B_r^2}{B^2}+\frac{4B_r}{rB} \right), \\
 p_{\parallel}&=&\frac{1}{A^2}\left( -2\frac{B_{tt}}{B}-\frac{B_t^2}{B^2}+2\frac{A_t}{A}\frac{B_t}{B}\right)\nonumber \\
 &&+\frac{1}{B^2}\left( \frac{B_r^2}{B^2}+2\frac{A_r}{A}\frac{B_r}{B}+\frac{2}{r}\frac{A_r}{A}+\frac{2}{r}\frac{B_r}{B}\right),\label{314p} \\
 p_{\perp}&=&-\frac{2}{A^2}\frac{B_{tt}}{B}+2\frac{A_t}{A^3}\frac{B_{t}}{B}-\frac{1}{A^2}\frac{B_t^2}{B^2}+\frac{1}{r}\frac{A_r}{A}\frac{1}{B^2}\nonumber\\
&&+\frac{1}{r}\frac{B_r}{B^3}+\frac{A_{rr}}{A}\frac{1}{B^2}-\frac{B_r^2}{B^4}+\frac{B_{rr}}{B^3}, \label{14r}\\
 q&=&-\frac{2}{AB^2}\left(-\frac{B_{rt}}{B} +\frac{B_rB_t}{B^2}+\frac{A_r}{A}\frac{B_t}{B}\right),\label{314d} 
 \end{eqnarray}
 \end{subequations}
for the line element \eqref{9}.
The equations (\ref{14}) describe the  gravitational interactions in the  interior of a shear-free  spherically symmetric star with heat flux and anisotropic pressures.

The boundary of a  radiating star divides the spacetime into   interior  and  exterior regions. The interior spacetime \eqref{9}  has to match across the boundary of the star to the Vaidya spacetime
\begin{eqnarray} 
ds^2&=& -\left(1-\frac{2m(v)}{R} \right)dv^2-2dvdR \nonumber \\
&&+R^2 \left( d\theta ^2+\sin ^2\theta d\phi^2 \right),  \label{va31}
\end{eqnarray}
which is the exterior. Here the quantity $m(v)$ denotes the mass of the star as measured by an observer at infinity. Matching    leads to the junction conditions
\begin{subequations}\label{rev31}
\begin{eqnarray}
Adt&=&\left[ \left(1-\frac{2m}{R_\Sigma}+2\frac{dR_\Sigma}{dv} \right)^\frac{1}{2}dv\right] _\Sigma ,\\
\left(rB \right)_{\Sigma}&=&R_{\Sigma },\\
m(v)&=&\left[\frac{r^3}{2}\left( \frac{BB_t^2}{A^2}-\frac{B_r^2}{B}\right)  -r^2B_r\right] _{\Sigma}, \\
(p_{\parallel})_{\Sigma}&=&(Bq)_{\Sigma},\label{santos1}
\end{eqnarray}
\end{subequations}
where  the hypersurface  $\Sigma $ defines the boundary of the radiating sphere. The particular junction condition \eqref{santos1} should be solved, together with  the field equations \eqref{14}, to obtain the potentials $A$ and $B$. This completes the model of a relativistic radiating star. The junction condition is a nonlinear differential equation  
\begin{eqnarray}\label{317}
& &2\frac{B_{rt}}{AB^2}+2\frac{B_{tt}}{A^2B}-2\frac{A_tB_t}{A^3B}-2\frac{B_rB_t}{AB^3}
-2\frac{A_rB_r}{AB^3}   \nonumber\\
& & -2\frac{A_rB_t}{A^2B^2}-\frac{B_r^2}{B^4} +\frac{B_t^2}{A^2B^2}-2\frac{A_r}{rAB^2}-2\frac{B_r}{rB^3}=0,
\end{eqnarray}
valid at the boundary of shear-free radiating star. Equation \eqref{317} is the  master equation that governs the evolution of the model. 
\section{The master equation \label{si3}}
We can use the Lie analysis of differential equations to find a solution to the master equation \eqref{317}. In a general relativistic context, the Lie analysis has proved useful in cosmological settings as shown by  Msomi \emph{et al.} \cite{b4,b5}, Kweyama \emph{et al.} \cite{b7,b77} and Nyonyi \emph{et al.} \cite{y1,y2}, and in astrophysical applications as illustrated in the treatments by Abebe \emph{et al.} \cite{b11,b12}. The general approach is described in these treatments and we will not repeat details here. Essentially we have to find  infinitesimal  Lie point generators that allow us to reduce a partial differential equation to ordinary differential equations. This  is made possible by the existence  of invariants associated with the Lie point symmetries of the partial differential equation equation \eqref{317}.

The process has been adapted over time and is now  algorithmic, and  so can be implemented by different computer software  packages. Utilizing the program  PROGRAM LIE  \cite{c13}, we  find that \eqref{317} admits the following
Lie point symmetries:
\begin{subequations}\label{318}
\begin{eqnarray}
G_1&=&-Af'(t)\frac{\partial }{\partial A}+f(t)\frac{\partial}{\partial t},\\
G_2&=&A\frac{\partial}{\partial A}+B\frac{\partial}{\partial B}, \\
G_3&=&A\frac{\partial}{\partial A}+r\frac{\partial}{\partial r},
\end{eqnarray}
\end{subequations}
where $f(t)$ is an arbitrary function of $t$. These symmetries generate invariants that can be used to reduce the partial differential equation \eqref{317} into ordinary differential equations for further analysis.
Using the symmetries in \eqref{318} in turn, or taking any linear combination, may be helpful in reducing the master equation into ordinary differential equations. Since the symmetries do not have a  nonzero Lie bracket relationship we do not consider the optimal system in this paper.  As   all  combinations of  these symmetries can be contained in a general linear combination, we take  
\begin{eqnarray}
aG_1+bG_2+cG_3&=&\left[ c+b-af'(t)\right] A\frac{\partial}{\partial A}+bB\frac{\partial}{\partial B} \nonumber\\
&& +af(t)\frac{\partial}{\partial t}+cr\frac{\partial}{\partial r}, \label{eto1}
\end{eqnarray}
 where $a$, $b$ and $c$ are arbitrary constants.
This combination  gives the invariants  
\begin{subequations}\label{inv3333}
\begin{eqnarray}
x&=&\frac{\exp\left( \int^t \frac{dt}{a f(t)} \right)}{r^{1/c}}, \\
A&=& \frac{h(x)}{f(t)} \exp \left(\int ^t \frac{cdt}{a f(t)}+\int ^t \frac{bdt}{a f(t)} \right) ,\\
B&=&g(x)r^{b/c},
\end{eqnarray}
\end{subequations}
 where $a\ne0$ and $c\ne0$ . The quantities $g(x)$ and $h(x)$ are arbitrary functions associated with the Lie symmetry generators and arise from integration. 
 
 %\section{ Class I: $a\ne 0$, $c\ne0$}
  Using the invariants \eqref{inv3333} we  can  write \eqref{317} in the form 
\begin{eqnarray} 
&&\left[2 a^2 g x^{2 b+2 c+1} \left((b+c) g-x g'\right)\right] h'h^2 \nonumber\\
&& +2 a c  x^{b+c+2} g^2 g' h'h-2 c^2 x^2 g^3 g' h'\nonumber\\
&&+ \left[c^2 x g^2 \left(x g'^2-2 g \left((b+c-1) g'-x g''\right)\right)\right]h \nonumber\\
&& - \left[2 a c g x^{b+c+1} \left(g \left(x g''+g'\right)-x g'^2\right)\right]h^2\nonumber\\
&&+\left[a^2 x^{2 (b+c)} \left(x g'-b g\right) \left((b+2 c) g-x g'\right)\right]h^3 =0,  \label{rrr}
\end{eqnarray}
where  primes denote differentiation with respect to  to the new variable $x$. Note that the partial differential equation \eqref{317} has been reduced to the ordinary differential equation \eqref{rrr}. Equation \eqref{rrr} is difficult to solve in general. To demonstrate an exact solution it  is necessary to make assumptions on the parameters and arbitrary functions.

 To progress we make the assumption 
 \begin{equation} g(x)=kh(x).
 \end{equation}
 Then  equation  \eqref{rrr} reduces to the form
\begin{eqnarray} 
&&2c k x^2  \left[c k-a x^{b+c}\right] h h''  \nonumber \\
&& -x^2  \left[3 a^2 x^{2 (b+c)}-4 a c k x^{b+c}+c^2 k^2\right]h'^2\nonumber\\
&&-2x  \left[c^3 k^2-2 a^2 b x^{2 (b+c)}-a c x^{b+c} \left(2 a x^{b+c}-k\right) \right. \nonumber\\
&& \left. +(b-1) c^2 k^2\right]h h' - a^2 b [b+2 c]x^{2 (b+c)} h^2=0.  \label{rrr1}
\end{eqnarray}
Equation \eqref{rrr1} is a  nonlinear second order  differential equation. We can reduce \eqref{rrr1} to first order if we let
\begin{equation}\label{gola} y=\frac{h'}{h}.
 \end{equation}
Then the transformation \eqref{gola} enables us to write \eqref{rrr1} in the form 
\begin{eqnarray}\label{rica10}
&& y'  + \left(\frac{3 a x^{b+c}}{2 c k}+\frac{1}{2}\right)y^2  +\left[  2 a^2 b x^{2 (b+c)}+a c x^{b+c} \right.\nonumber\\
&&\times\left.\left(2 a x^{b+c}-k\right)+(1-b) c^2 k^2-c^3 k^2\right]\nonumber\\
&& \left[ c k x \left(c k-a x^{b+c}\right)\right] ^{-1}y +\frac{a^2 b (b+2 c) x^{2 (b+c-1)}}{2 c k \left(a x^{b+c}-c k\right)}=0.\nonumber\\
&&
 \end{eqnarray}
 Observe that \eqref{rica10} is a Riccati equation in the variable $y$. Riccati equations can be transformed  to second order linear equations. We let
 \begin{equation}\label{fu1}
u(x)=\exp\left[ \int^x\left(\frac{3 a x^{b+c}}{2 c k}+\frac{1}{2} \right)  y(x)dx\right].
\end{equation}
Note that in equation \eqref{fu1} the term $\left(\frac{3 a x^{b+c}}{2 c k}+\frac{1}{2} \right)$ is the coefficient of the quantity $y^2$ in \eqref{rica10}. Using \eqref{fu1} equation \eqref{rica10} is transformed to 
\begin{equation} \label{line}
 u''+\gamma(x)u' +\zeta(x)u=0,\qquad
\end{equation}
where 
\begin{subequations}
\begin{eqnarray} \gamma(x)&=&\left[ 6 a^3 b x^{3 (b+c)}+a^2 c x^{2 (b+c)} \left(6 a x^{b+c}+(5 b-3) k\right) \right. \nonumber \\
&& \left. -c^3 k^2 \left(6 a x^{b+c}+(b-1) k\right)\right. \nonumber\\
&&\left. + a c^2 k x^{b+c} \left(5 a x^{b+c}+(2-6 b) k\right)-c^4 k^3\right] \nonumber\\
&&\times \left[c k x \left(-3 a^2 x^{2 (b+c)}+2 a c k x^{b+c}+c^2 k^2\right) \right] ^{-1},\nonumber\\
&&\\
\zeta(x)&=&\frac{a^2 b (b+2 c) x^{2 (b+c-1)} \left(3 a x^{b+c}+c k\right)}{4 c^2 k^2 \left(a x^{b+c}-c k\right)}.
\end{eqnarray}
\end{subequations}
Therefore we have the remarkable feature that the second order nonlinear equation \eqref{rrr1} has been transformed to the linear equation \eqref{line} via the transformations
\eqref{gola} and \eqref{fu1}. 

Particular choices of the constants  $a,b$ and $c$ allow us to integrate \eqref{rrr}, \eqref{rrr1} or \eqref{line} and obtain potentials for the gravitational field. There are several classes that arise which  we consider in turn.
\section{Class I \label{sec4}}
In this class we make the restrictions that $a\ne0$ and $c\ne0$. These conditions arise so that the invariants \eqref{inv3333} exist.
\subsection{Case I(a): $b=0$} If we set $b=0$ then the coefficient  of the linear term in $u$ in  \eqref{line}  disappears. In this case we have  
\begin{eqnarray}
&& u''+ \left[ 3 a^2 x^{2 c} \left(2 a x^c-k\right)+c^2 k^2 \left(k-6 a x^c\right)\right.  \nonumber \\
&&\left. +a c k x^c \left(5 a x^c+2 k\right)-c^3 k^3\right] \left[ k x \left(-3 a^2 x^{2 c}\right. \right.  \nonumber \\
&&\left.\left. +2 a c k x^c+c^2 k^2\right)\right] ^{-1} u'=0,
\end{eqnarray}
which is a separable equation.
This  can be easily integrated to give
\begin{equation}
u(x)=d \exp\left( \frac{2 a x^c}{c k}\right) \left(6 a^2 x^{2 c}-10 a c k x^c+3 c^2 k^2\right)+\tilde{m},
\end{equation}
where $d$ and $\tilde{m}$ are constants.
 Then equation \eqref{fu1} yields
\begin{equation}
y(x)=\frac{8 a c x^{c-1} \exp\left( \frac{2 a x^c}{c k}\right)  \left(a x^c-c k\right)}{\exp\left( \frac{2 a x^c}{c k}\right) \left( 6 a^2 x^{2 c} +3 c^2  k^2 -10 a c  k x^c \right) + m},
\end{equation}
where $m =\frac{\tilde{m}}{d}$ is a new constant.

Hence  we obtain the potentials
\begin{subequations}\label{sss1}
\begin{eqnarray}
A&=& \frac{n  }{f(t)} \exp \left(\int^x\left[ 8 a cx^{c-1} \exp\left( \frac{2 ax^c}{c k}\right) \left(ax^c-c k\right)\right]\right. \nonumber\\
&&\times\left.\left[  \exp\left( \frac{2 ax^c}{c k}\right) \left( 6 a^2x^{2 c}  - 10a c kx^c \right.\right.\right. \nonumber\\
&&\left.\left.\left.+3 c^2 k^2\right) + m \right] ^{-1}\, dx+\int^t \frac{cdt}{a f(t)} \right), \label{ffuu1s} \\
B&=&\frac{k f(t)}{\exp \left(\int ^t\frac{cdt}{a f(t)} \right)}A,
\end{eqnarray}
\end{subequations}
where $x=\exp\left( \int ^t\frac{\, dt}{af(t)} \right) r^{-1/c}$ and $n$ is an arbitrary constant of integration. We believe that this solution is not contained in the literature. 

It is possible to evaluate the integral containing $x$ in \eqref{ffuu1s} when  $m=0$. In this case 
\begin{eqnarray}
A&=&\tilde{n}  \left(6 a^2 x^{2 c}-10 a c k x^c+3 c^2 k^2\right)^{2/3}\nonumber\\
&&\times \frac{ \exp \left(\int^t \frac{cdt}{a f(t)} \right)}{f(t)} \left(\frac{7-\frac{3 \sqrt{7} c k x^{-c}}{a}+5 \sqrt{7}}{\frac{3 \sqrt{7} c k x^{-c}}{a}-5 \sqrt{7}+7}\right)^{\frac{4}{\sqrt{7}}}\nonumber\\
&&\times\left(\frac{7-\frac{6 \sqrt{7} a x^c}{c k}+5 \sqrt{7}}{\frac{6 \sqrt{7} a x^c}{c k}-5 \sqrt{7}+7}\right)^{\frac{10}{3 \sqrt{7}}},
\end{eqnarray}
and $\tilde{n}$ is a new constant. For this special case the explicit dependence on the variable $x$ is fully specified; there is freedom only in the variable $t$.

\subsection{Case I(b): $b=-c$ }
If  we  set $ b=-c$  then equation \eqref{line} becomes
\begin{equation} \label{fu2}
x^2u''+xu'+\frac{a^2 (3 a+c k)}{4 k^2 (c k-a)}u=0,
\end{equation}
  which is a simpler form. It is interesting to note that this case produces the Euler equation \eqref{fu2}. We can integrate \eqref{fu2} to  obtain
\begin{eqnarray}
u(x)&=&\tilde{c_1 }\cosh \left(\frac{a  \sqrt{3 a+c k}}{2 k \sqrt{a-c k}}\log (x)\right)\nonumber\\
&&+ \tilde{c_2} \sinh \left(\frac{a  \sqrt{3 a+c k}}{2 k \sqrt{a-c k}}\log (x)\right),
\end{eqnarray}
where $\tilde{c_1 }$ and $\tilde{c_2}$ are arbitrary constants of integration. Then from \eqref{fu1} we obtain 
\begin{equation}
y(x)=\frac{a c \sqrt{3 a+c k} }{\sqrt{a-c k} }\frac{ c_1x^{\frac{a \sqrt{3 a+c k}}{k \sqrt{a-c k}}}-c_2}{ x \left(3 a x^c+c k\right) \left(c_1 x^{\frac{a \sqrt{3 a+c k}}{k \sqrt{a-c k}}}+c_2\right)},
\end{equation}
where $c_1=\tilde{c_1}+\tilde{c_2}$ and $c_2=\tilde{c_1}-\tilde{c_2}$.

Hence we have the potentials
\begin{subequations}\label{rtt}
\begin{eqnarray}
A&=& \frac{1}{f(t)} \left(m \left[ r^{-1/c}\exp\left( \int^t \frac{dt}{a f(t)} \right)\right]^{\frac{a \sqrt{3 a+c k}}{2 k \sqrt{a-c k}}}\right. \nonumber\\
&&\left. +n \left[ r^{-1/c}\exp\left( \int ^t\frac{dt}{a f(t)} \right)\right]^{-\frac{a \sqrt{3 a+c k}}{2 k \sqrt{a-c k}}}\right){}^{\frac{2 c k}{3 a+c k}},\nonumber \\
&&\\
B&=& k\frac{f(t)}{r} A,
\end{eqnarray}
\end{subequations}
 where $m=c_1c_3^{\frac{3a+ck}{2ck}}$ and $n=c_2c_3^{\frac{3a+ck}{2ck}}$ are  constants. This is  another new solution to  the master equation.
\subsection{Subclass I(c): $b=-c$, $k=-3a/c$ }
Note that in \eqref{rtt}, $k\ne-3a/c$. With the values  $b=-c$ and  $k=-3a/c$ the transformation \eqref{fu1} leads to an inconsistency. This means that we have to integrate \eqref{rrr1} or \eqref{rica10} for this case. If we set $k=-3a/c$ and $b=-c$, then  equation \eqref{rrr1} becomes 
\begin{equation} \label{27}
24 x^2 h h''-24 x^2 h'^2+24 x h h'+c^2 h^2=0,
\end{equation}
which  is greatly simplified. Now \eqref{27}  can be integrated to give
\begin{equation}
h(x)= \frac{n x^{m}}{\exp\left[ \frac{c^2}{48}  \log ^2(x)\right] },
\end{equation}
where $m$ and $n$ are constants of integration.

Hence we get the metric functions
\begin{subequations}\label{210}
\begin{eqnarray}
A&=& \frac{n}{f(t)}  \frac{  \left[ r^{-1/c}\exp\left( \int^t \frac{dt}{a f(t)} \right)\right]^{m}}{\exp\left( \frac{c^2}{48}  \log ^2 \left[ r^{-1/c}\exp\left( \int^t \frac{dt}{a f(t)} \right)\right]\right) }, \\
B&=&-\frac{3a}{c}\frac{f(t)}{r} A.
\end{eqnarray}
\end{subequations} 
This is also another new solution  to the master equation \eqref{317}.

\subsection{Case I(d): $b=-2c$  }
If we set $b=-2c$ then the coefficient of the linear term in $u$ in  \eqref{line}  disappears as in Case I(a). Then  we obtain 
\begin{eqnarray}
&&u''-\left[  \left(x^{-c-1} \left(6 a^3+a^2 (5 c+3) k x^c-2 a c (3 c+1) k^2 x^{2 c}\right.\right. \right.  \nonumber\\
&&\left.\left. \left.  -c^2 (c+1) k^3 x^{3 c}\right)\right)\right] \left[ k \left(3 a^2-2 a c k x^c\right. \right.  \nonumber\\
&&\left.\left. -c^2 k^2 x^{2 c}\right)\right] ^{-1}u'=0,
\end{eqnarray}
which is separable. This  can be easily integrated to give
\begin{eqnarray}
u(x)&=&d k \exp \left( \frac{2 a x^{-c}}{c k}\right)  \left(\frac{3 c^2 k^2}{a}+6 a x^{-2 c}\right.  \nonumber\\
&&\left.-10 c k x^{-c}\right)+\tilde{m},
\end{eqnarray}
where the constants  $d$ and $\tilde{m} $ result from the integration. Then equation \eqref{fu1} yields
\begin{eqnarray}
y(x)&=&\left[ 8 a c \exp\left( \frac{2 a x^{-c}}{c k}\right) \left(c k x^c-a\right)\right] \nonumber\\
&&\times\left[ x \left(\exp\left( \frac{2 a x^{-c}}{c k}\right)  \left[ 6 a^2 +3 c^2 k^2 x^{2 c} \right.\right.\right.   \nonumber\\
&&\left.\left.\left.-10 a c k x^c \right] +m x^{2 c}\right)\right] ^{-1},
\end{eqnarray}
where $m =\frac{\tilde{m}}{d}$ is a  constant.

Hence  we  can generate the potentials
\begin{subequations}\label{sss}
\begin{eqnarray}
A&=&\frac{n }{f(t)} \exp \left(\int^x \left[ 8 a c \exp\left( \frac{2 a x^{-c}}{c k}\right) \left(c k x^c-a\right)\right] \right.   \nonumber\\
&&\left.  \times\left[ x \left(\exp\left( \frac{2 a x^{-c}}{c k}\right)  \left[ 6 a^2 +3 c^2 k^2 x^{2 c} -10 a c k x^c \right]\right. \right.\right.   \nonumber\\
&&\left.\left. \left. +m x^{2 c}\right)\right] ^{-1}\, dx-\int^t \frac{cdt}{a f(t)}\right),  \\
B&=&\frac{k f(t)}{\exp \left(\int^t \frac{cdt}{a f(t)} \right)}A,
\end{eqnarray}
\end{subequations}
where $x=\exp\left( \int^t \frac{\, dt}{af(t)} \right) r^{-1/c}$ and $n$ is new arbitrary constant. This is a new model.

This class of solution contains a previously found model. To show this  we set 
\begin{equation}
 a=c=k=1 \text{ and } f(t)=t.
\end{equation}
Then  solution \eqref{sss} becomes
 \begin{eqnarray}
A&=&B\nonumber\\
&=&d\exp \left(\int_1^{t/r} \frac{8\exp\left( 2z\right)  (z-1)}{\exp\left( 2z\right) \left( 6 z^2 -10 z  +3 \right) +m} \, dz\right),\nonumber\\
&&
\end{eqnarray}
which is a self-similar solution for the master equation. It has been obtained previously by Abebe \emph{et al.} \cite{b10} when analyzing  a    conformally flat radiating star.

\section{Class II \label{sec5}} 
 In this class we set $a=0$. Then the invariants \eqref{inv3333} do not exist. For this category of solution we then  have to consider the symmetry 
  \begin{equation}
   \label{sodo} bG_2+cG_3=\left( c+b\right) A\frac{\partial}{\partial A}+bB\frac{\partial}{\partial B}+cr\frac{\partial}{\partial r},
  \end{equation}
from \eqref{eto1}. The invariants associated with \eqref{sodo} are \begin{subequations}\label{si1q}
 \begin{eqnarray}
A&=&h(t)r^{(b+c)/c},\\
B&=&g(t)r^{b/c}, \text{ and }\\
t.&&
\end{eqnarray}
 \end{subequations}
The invariants \eqref{si1q} reduce the master equation \eqref{317} to
\begin{eqnarray} \label{si2}
&&2 c^2 g g' h'-c^2 \left(2 g g''+g'^2\right)h+2 c (b+c)  g'h^2\nonumber\\
&&+\left(3 b^2+6 b c+2 c^2\right) h^3=0.
\end{eqnarray}
Here the primes stand for the derivatives with respect to the independent variable $t$. Equation \eqref{si2} is an Abel differential equation in $h$. It is not possible to integrate it in general. Particular solutions  do exist as  we now demonstrate.  We set 
\begin{equation} b= \left(\pm\frac{\sqrt{3}}{3} -1\right)c,
\end{equation}
to simplify the Abel equation. Then \eqref{si2} becomes 
\begin{equation} \label{ffuu}
6 g g'h'-3  \left(2 g g''+g'^2\right)h\pm 2 \sqrt{3}  g'h^2=0.
\end{equation}
The advantage of \eqref{ffuu} is that it is a Bernoulli equation in $h$. It is interesting to note that it  can be integrated even when the function $g$ is unspecified. We integrate \eqref{ffuu} to obtain 
\begin{equation}
h(t)=\pm\frac{\sqrt{3} }{2  }\frac{g'\sqrt{g} }{\sqrt{g}+ d},
\end{equation}
where $d$ is a constant of integration.

Hence the potentials functions become
 \begin{subequations}\label{si1}
\begin{eqnarray}
A&=&\pm\frac{\sqrt{3} }{2  }\frac{g'\sqrt{g} }{\sqrt{g}+ d}r^{ \pm\frac{\sqrt{3}}{3} },\\
B&=&gr^{ \pm\frac{\sqrt{3}}{3} -1 },
\end{eqnarray}
\end{subequations} 
which is a new exact solution for the shear-free model.

Other exact solutions to \eqref{si2} exist but they may not be realistic. For example, \eqref{si2}  can be integrated to yield exact solutions if we assume $c=-b$ and $g(t)=kh(t)$. Unfortunately the model then become unphysical as the heat flux vanishes  when  $g(t)=k h(t)$ and both the tangential pressure and the energy density vanish when $c=-b$.

\section{Class III \label{sec6}}
In this class we set $c=0$. Consequently the invariants \eqref{inv3333} are not defined. Therefore in this case we consider the symmetry 
\begin{equation}
 \label{tame} aG_1+bG_2=(b-af'(t))A\frac{\partial}{\partial A}+B\frac{\partial}{\partial B}+af(t)\frac{\partial}{\partial t},
 \end{equation} 
 from \eqref{eto1}. Note that the invariants arising from \eqref{tame} are given by 
\begin{subequations} \label{tigire}
\begin{eqnarray} 
A&=&h(r)\frac{\exp\left(\int \frac{bdt}{af(t)} \right) }{f(t)}, \\
B&=&g(r)\exp\left(\int \frac{bdt}{af(t)}\right),\text{ and }\\
r.&&
\end{eqnarray}
\end{subequations} 
Using the invariants \eqref{tigire} the master equation \eqref{317} is reduced to give  
\begin{eqnarray} \label{aro}
&&2 a^2 g  \left(r g'+g\right)h h'+2 a b r g^3 h'+a^2  g' \left(r g'+2 g\right)h^2\nonumber\\
&&-b^2 r g^4=0,
\end{eqnarray}
where the primes denote derivatives with respect to the independent variable $r$. Equation \eqref{aro} is highly nonlinear and cannot be easily integrated  in general. We assume that \begin{equation}  g(r)= k h(r),
\end{equation}
so that  \eqref{aro} becomes
\begin{equation}\label{lebba} 
\left(\frac{h'}{h}\right)^2+\frac{ 2}{3 } \left(\frac{b k}{a}+\frac{2}{r}\right)\frac{h'}{h} -\frac{b^2 k^2}{3 a^2}=0.
\end{equation}
We observe that \eqref{lebba} may be treated as a quadratic equation  in the quantity $\frac{h'}{h}$; this gives
\begin{equation} \label{lebba2}
\frac{h'}{h}=\frac{\pm2 \sqrt{a^2 \left(a^2+a b k r+b^2 k^2 r^2\right)}-2 a^2-a b k r}{3 a^2 r}.
\end{equation}
The two roots in \eqref{lebba2} can be integrated to yield  two different classes of solutions
\begin{subequations}
\begin{eqnarray}
h(r)&=& \frac{d}{r^{4/3}}\left( \frac{\left( \alpha (r)+2 a^2+a b k r\right)^{2} }{ \left(\alpha (r)+2 a b k r+a^2\right) }\right)^{1/3} \nonumber\\
&&\times\exp \left(\frac{ -\alpha (r)-ab k r}{3 a^2}\right),\\
h(r)&=&d\left( \frac{\alpha (r)+2ab k r+a^2 }{\left( \alpha (r)+2 a^2+a b k r\right)^{2}}\right)^{1/3} 
\nonumber\\
&&\times\exp \left(\frac{ \alpha (r)-a b k r}{3 a^2}\right),
\end{eqnarray}
\end{subequations}
where $\alpha (r)= 2 a\sqrt{a^2+a b k r+b^2 k^2 r^2}$ and $d$ is an arbitrary constant of integration.

Hence we have
\begin{subequations}\label{inv333s2}
\begin{eqnarray}
A&=&\frac{d}{r^{4/3}f(t)}\left( \frac{\left( \alpha (r)+2 a^2+a b k r\right)^{2} }{ \left(\alpha (r)+2a b k r+a^2\right) }\right)^{1/3}\nonumber\\
&&\times \exp \left(\int \frac{bdt}{af(t)} -\frac{ \alpha (r)}{3 a^2}-\frac{b k r}{3 a}\right),\\
B&=&kf(t) A,
\end{eqnarray}
\end{subequations}
 and 
\begin{subequations}\label{inv3332}
\begin{eqnarray}
A&=&\frac{d}{f(t)}\left( \frac{\alpha (r)+2 a b k r+a^2 }{\left( \alpha (r)+2 a^2+a b k r\right)^{2}}\right)^{1/3} \nonumber\\
&&\times\exp \left( \int \frac{bdt}{af(t)}+\frac{\alpha (r)}{3 a^2}-\frac{b k r}{3 a}\right),\\
B&=&kf(t)A ,
\end{eqnarray}
\end{subequations}
 which are  two classes of solution to the master equation.
 
 This category of solutions  reduces to known models. We can show this by setting 
\begin{equation} 
f(t)=1 \text{ and } b=d=k=1.
\end{equation}
Then the  solutions \eqref{inv333s2} and \eqref{inv3332} become
\begin{subequations}
\begin{eqnarray}
A&=&B\nonumber\\
&=&r^{-4/3}\exp\left(  \frac{t}{a}-\frac{\alpha(r)}{3 a}-\frac{r}{3 a}\right)\nonumber\\
&&\times\left( \frac{\left(\alpha(r)+2 a+r\right)^{2}  }{ \left(  \alpha(r)+a+2 r\right) }\right) ^{1/3},\\
A&=&B \nonumber\\
&=&\exp\left( \frac{t}{a}- \frac{r}{3 a}+ \frac{\alpha(r)}{3 a}\right)\nonumber\\
&&\times\left( \frac{\alpha(r)+a+2 r }{ \left(\alpha(r)+2 a+ r\right)^{2}}\right) ^{1/3},
\end{eqnarray}
\end{subequations}
respectively. These solutions were previously obtained by Abebe \emph{et al.} \cite{b10} for a  radiating star which has the  property of conformal flatness.

\section{Equations of state\label{s5}}
In relativistic astrophysics it is important that the model should admit an equation of state on physical grounds. Many exact solutions for a radiating star that have been found  before do not satisfy this condition. We can report that particular  classes of models in this paper do admit a linear equation of state which are barotropic. We give each of these cases below and associate them with the relevant Lie symmetry.
\subsection*{(a) The generator $aG_1-cG_2+cG_3$ $(k\neq -\frac{3a}{c})$ }
 
Using the generator  $aG_1-cG_2+cG_3$ we can obtain a model from \eqref{rtt} which admits an equation of state. The line element for this  case is
\begin{eqnarray}\label{eto}
 ds^2&=&\left[  \left(m\psi ^{1/2}+n \psi ^{-1/2}\right){}^{\frac{2 c k}{3 a+c k}}\right] ^2\left(-\left[ \frac{1}{f(t)}\right] ^2 dt^2\right. \nonumber\\
&&\left. +\left[\frac{k}{r}  \right] ^2\left[  dr^2+r^2\left( d\theta^2+\sin ^2\theta d\phi^2 \right) \right]\right) , 
 \end{eqnarray}
where $\psi=\left[r^{-1/c} \exp \left( \int \frac{\, dt}{af(t)} \right) \right]^{\frac{a \sqrt{3 a+c k}}{ k \sqrt{a-c k}}}$. The matter variables become
\begin{subequations}
 \begin{eqnarray}
 \mu&=& \left[2 \left(a^2-a c k+c^2 k^2\right)  \left(c k \left(m^2     \psi ^{2}-4 m n \psi+n^2\right)\right.\right.  \nonumber\\
&&\left.\left. -6 a m n \psi\right)\right] \left[ c k^3 (a-c k) (3 a+c k) \left(m \psi+n\right)^2\right.  \nonumber\\
&&\left.\times\left(\psi ^{-1/2}\left(m \psi+n\right)\right)^{\frac{4 c k}{3 a+c k}}\right]^{-1}  ,\\
 p_{\parallel} &=&\left[ 2 a  \left(6 a m n \psi-c k \left(m^2\psi ^{2}-4 m n \psi+n^2\right)\right)\right]  \nonumber\\
&&\times\left[ k^2 (c k-a) (3 a+c k) \left(m \psi+n\right)^2\right.  \nonumber\\
&&\left.\times\left(\psi ^{-1/2}\left(m \psi+n\right)\right)^{\frac{4 c k}{3 a+c k}}\right] ^{-1},\\
p_{\perp}&=&\left[ (a+c k) \left(\psi ^{-1/2}\left(m \psi+n\right)\right)^{-\frac{4 c k}{3 a+c k}} \left(12 a m n \psi\right.\right.   \nonumber\\
&&\left.\left. +c k \left(m \psi+n\right)^2\right)\right] \left[ c k^3 (3 a+c k) \left(m \psi+n\right)^2\right.  \nonumber\\
&&\left.\times\left(\psi ^{-1/2}\left(m \psi+n\right)\right)^{\frac{4 c k}{3 a+c k}}\right] ^{-1},\\
 q&=&\left[ \frac{k}{r}\left(m\psi ^{1/2}+n \psi ^{-1/2}\right){}^{\frac{2 c k}{3 a+c k}}\right] ^{-1} p_{\parallel}.
 \end{eqnarray}
 \end{subequations}
 From the above we generate the linear barotropic equation of state
\begin{equation}
p_ \parallel=\lambda\mu, \quad \lambda =\frac{a c k}{a^2-a c k+c^2 k^2},
\end{equation}
 provided that $k\neq-\frac{3a}{c}$.

\subsection*{(b) The generator $aG_1-cG_2+cG_3$ $(k= -\frac{3a}{c})$ }

The generator $aG_1-cG_2+cG_3$ helps to find another model  \eqref{210}  when $k=-\frac{3a}{c}$. This model also admits  an equation of state. The line element for  this case is given by
\begin{eqnarray} \label{drogba}
ds^2&=&\left[  \frac{  n \varphi^{m}}{\exp\left( \frac{c^2}{48}  \log ^2 \varphi\right) } \right] ^2\left[ -\left( 1/f(t)\right) ^2dt^2+9\left( a/(cr)\right) ^2\right.  \nonumber\\
&&\left.\times\left[  dr^2+r^2\left( d\theta^2+\sin ^2\theta d\phi^2 \right) \right] \right],
\end{eqnarray}
where $ \varphi= r^{-1/c}\exp\left( \int \frac{dt}{a f(t)} \right)$.
The matter variables become
\begin{subequations}
 \begin{eqnarray}
 \mu&=&\left[13  \left(c^4 \log ^2\varphi-48 m c^2 \log \varphi+24 \left(c^2+24 m^2\right)\right) \right.\nonumber\\
 &&\left.\times\exp \left(\frac{1}{24} c^2 \log ^2\varphi\right)\right] \left[ 2592 a^2 n^2\varphi^{2 m}\right] ^{-1},\\
 p_{\parallel} &=&\left[  \left(48 m c^2 \log \varphi-c^4 \log ^2\varphi-24 \left(c^2+24 m^2\right)\right) \right.\nonumber\\
 &&\left.\times\exp \left(\frac{1}{24} c^2 \log ^2\varphi\right)\right] \left[ 864 a^2 n^2\varphi^{2 m}\right] ^{-1},\\
p_{\perp}&=&\left[ \left(48 m c^2 \log \varphi-c^4 \log ^2\varphi+48 c^2+576 m^2\right)\right.\nonumber\\
 &&\left.\times \exp \left(\frac{1}{24} c^2 \log ^2\varphi\right)\right] \left[ 648 a^2 n^2\varphi^{2 m}\right] ^{-1},\\
 q&=& \left[ \frac{3 a}{cr}\frac{  n \varphi^{m}}{\exp\left( \frac{c^2}{48}  \log ^2 \varphi\right) }\right] ^{-1}p_{\parallel}.
 \end{eqnarray}
 \end{subequations}
 We observe that  this case also  admits an equation of state
\begin{equation}
p_ \parallel=\lambda\mu, \quad \lambda =-\frac{3}{13},
\end{equation}
which is linear and barotropic.
\subsection*{(c) The generator $bG_2+cG_3$ }
The generator $bG_2+cG_3$ can also associated with an equation of state. The line element for this model becomes 
\begin{eqnarray}\label{iis1}
 ds^2&=&-\frac{3}{4}g'^2r^{ \pm2\frac{\sqrt{3}}{3} } dt^2+g^2r^{ \pm\frac{2\sqrt{3}}{3} -2 } \left[  dr^2\right. \nonumber\\
&&\left.+r^2\left( d\theta^2+\sin ^2\theta d\phi^2 \right) \right]. 
 \end{eqnarray}
from \eqref{si1}. In the above we have set  the arbitrary constant $d=0$ without any loss of any generality. 
 The matter variables become
 \begin{subequations}
\begin{eqnarray}
\mu&=&  \frac{14 r^{\mp\frac{2}{\sqrt{3}}}}{3 g^2}  ,\\ 
 p_ \parallel&=& -\frac{4 r^{\mp\frac{2}{\sqrt{3}}}}{3 g^2} , \\ p_{\perp}&=&  -\frac{ r^{\mp\frac{2}{\sqrt{3}}}}{g^2} ,\\q&=&\left[gr^{ \pm\frac{\sqrt{3}}{3} -1 } \right] ^{-1}p_ \parallel.
\end{eqnarray}
\end{subequations}
 This solution  also satisfies the barotropic equation of state
\begin{equation}
p_ \parallel=\lambda\mu, \quad \lambda =-\frac{2}{7},
\end{equation}
 which is linear.

\section{Discussion \label{s7}}
Our treatment indicates an interesting interplay between Lie symmetries, Lie algebras, nonlinear differential equations and a 
radiating star in relativistic astrophysics.
We have analyzed a radiating star with a shear-free matter distribution with anisotropic stress using the Lie analysis of differential equations. A systematic study of the master equation, governing the evolution of the radiating star, was undertaken. Three classes of new exact solutions were generated. In each case the Lie infinitesimal generators can be identified and the gravitational potentials may be written explicitly. In certain cases we find that our models contain solutions found previously. In particular we regain the conformally flat solutions  of Abebe \emph{et al.} \cite{b10}. We also find that three exact models, for specific forms of the Lie point symmetry, contain  a linear equation of state. The line elements and the equations of state can be  written in explicit form. We summarize our results in Table \ref{table1} and Table \ref{table2}.  In Table \ref{table1} the Lie point symmetries and corresponding invariants are listed.  The restrictions  on the arbitrary functions $g(x)$ and $h(x)$ and the constants $a,b,c,$ and $k$ are given.  The resulting gravitational potentials $A(r,t)$ and $B(r,t)$ are written explicitly. 
 Table \ref{table2} identifies the symmetries with those particular exact solutions for which an explicit  linear barotropic equation of state exists.

\begin{figure} [T]
\begin{center}
\includegraphics[scale=.5]{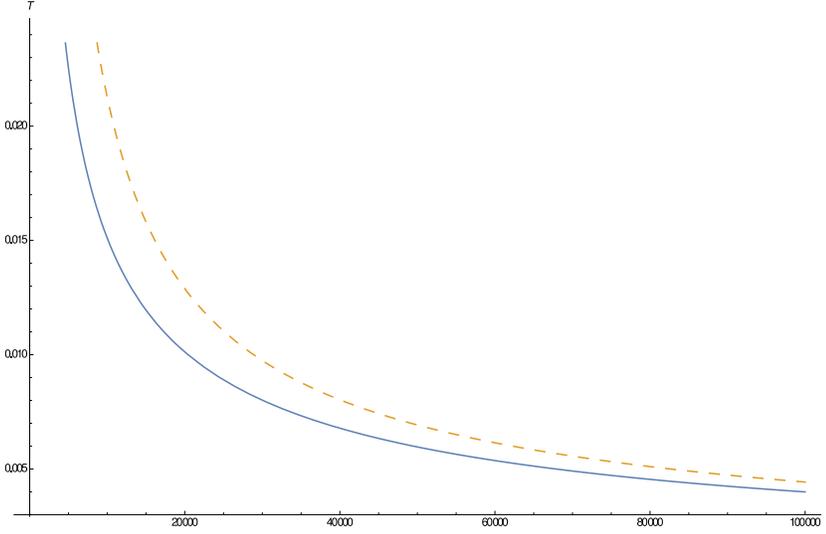}
\end{center}
\caption{Causal temperature (dashed line), noncausal temperature (solid line)   versus $r$\label{fig}}
\end{figure}

The solutions found in this paper may be used to study the physical properties of a relativistic radiating star. To illustrate this we consider the temperature profiles.
The causal  heat transport equation becomes 
\begin{equation} \label{taa}
\tau \left( qB\right) _t+ABq=-\frac{\kappa}{B}\left(AT \right) _r.
\end{equation}
The coefficient of thermal conductivity $\kappa $ and the relaxation time $\tau$ are chosen as 
\begin{equation} \label{kaa} 
\kappa = \gamma T^3 \tau _c, \qquad \tau _c=\left( \frac{\alpha}{\gamma}\right) T^{-\sigma}, \qquad \tau = \left(\frac{\beta\gamma}{\alpha} \right) \tau _c,
\end{equation}
based on physical grounds as shown by Govender and coauthors \cite{cc13, cd13}. In equation \eqref{kaa} $\alpha \geq 0, \beta \geq 0$, and $\gamma \geq 0$ are constants and $\tau _c$ is the mean collision time between massless and massive particles. We consider the special case of mean collision time $\sigma=0$ for simplicity. 
Then \eqref{taa} can be integrated to give  
\begin{equation}
\left( AT\right) ^4=-\frac{4}{\alpha}\left( \beta \int A^3B(qB)_tdr+\int A^4 qB^2dr\right)+w(t),
\end{equation}
which is the causal temperature. If we choose the gravitational potentials in  \eqref{si1}, the causal temperature can written  explicitly explicitly as 
 \begin{eqnarray}
 T^4 &=&\left(w(t)-\left[ r^{\frac{2}{\sqrt{3}}} g'(t)^4 \left(30 d \beta+\sqrt{3} \alpha r^{\frac{1}{\sqrt{3}}} g(t)^{3/2}\right.\right.\right. \nonumber\\
 &&\left.\left.\left. +24 \beta \sqrt{g(t)}\right)\right] \left[ 4 \alpha g(t) \left(d+\sqrt{g(t)}\right){}^3\right] ^{-1}\right)  \nonumber\\
&&\times\frac{16  \left(d+\sqrt{g(t)}\right){}^4}{9r^{\frac{4}{\sqrt{3}}} g(t)^2 g'(t)^4}.
 \end{eqnarray}
 Figure \ref{fig}  then gives the graphical behavior of the temperature  when $\alpha=d=-1$, $g(t)=t=0.5$ and $w(t)=-1$.
 
We find that the temperature is decreasing as we approach the boundary and the causal temperature is greater than the Eckart temperature. This is consistent with other treatments (see for example the recent analysis of Reddy \emph{et al.} \cite{bb101}). 

\begin{table*}[T]
\centering
\caption{Symmetry, invariants, restrictions and resulting gravitational potentials \label{table1} }
\begin{tabular*}{\textwidth}{@{\extracolsep{\fill}}llll@{}}
                                                                          
    \hline
Symmetry& Invariants       &Restrictions &Gravitational potentials  \\
&       && \\
   \hline 
      $aG_1+bG_2$  &$x=\frac{\exp\left( \int ^t \frac{dt}{a f(t)} \right)}{r^{1/c}},$&$g=kh, b=0$&$A= \frac{n  }{f(t)} \exp \left(\int^x\frac{8 a cx^{c-1} \exp\left( \frac{2 ax^c}{c k}\right) \left(ax^c-c k\right)}{\exp\left( \frac{2 ax^c}{c k}\right) \left( 6 a^2x^{2 c}  - 10a c kx^c +3 c^2 k^2\right) + m} \, dx+\int^t \frac{cdt}{a f(t)} \right),$\\
   
      $\qquad+cG_3$&$A= \frac{h(x)}{f(t)}$& &$B=\frac{k f(t)}{\exp \left(\int^t \frac{cdt}{a f(t)} \right)}A$ \qquad where $x=\exp\left( \int ^t \frac{\, dt}{af(t)} \right) r^{-1/c}$\\

  \cline{3-4}
   &$\times \exp \left(\int ^t \frac{cdt}{a f(t)}\right.  $&$g=kh, b=-2c$&$A= \frac{n }{f(t)} \exp \left(\int^x \frac{8 a c \exp\left( \frac{2 a x^{-c}}{c k}\right) \left(c k x^c-a\right)}{x \left(\exp\left( \frac{2 a x^{-c}}{c k}\right) \left[ 6 a^2 +3 c^2 k^2 x^{2 c} -10 a c k x^c \right] +m x^{2 c}\right)} \, dx\right.$\\
  
   &$\left. +\int^t \frac{bdt}{a f(t)} \right),$&&$\qquad\quad-\left.\int^t \frac{cdt}{a f(t)}\right),$\\
      &&&$B=\frac{k f(t)}{r^2\exp \left(\int ^t \frac{-cdt}{a f(t)} \right)}A,$ \qquad where $x=\exp\left( \int ^t \frac{\, dt}{af(t)} \right) r^{-1/c}$\\

  \cline{3-4}
     
     &$B=g(x)r^{b/c}$&$g=kh, b=-c$&   $A=\frac{1}{f(t)} \left(m \left[ r^{-1/c}\exp\left( \int ^t \frac{dt}{a f(t)} \right)\right]^{\frac{a \sqrt{3 a+c k}}{2 k \sqrt{a-c k}}}\right.$\\

    &&&$\qquad \left. +n \left[ r^{-1/c}\exp\left( \int ^t\frac{dt}{a f(t)} \right)\right]^{-\frac{a \sqrt{3 a+c k}}{2 k \sqrt{a-c k}}}\right){}^{\frac{2 c k}{3 a+c k}},$\\
   
&& &$B=\frac{kf(t)}{r} A$\\

  \cline{3-4}
     &&$g=kh, b=-c,$&$A=\frac{n}{f(t)}  \frac{  \left[ r^{-1/c}\exp\left( \int^t \frac{dt}{a f(t)} \right)\right]^{m}}{\exp\left( \frac{c^2}{48}  \log ^2 \left[ r^{-1/c}\exp\left( \int ^t \frac{dt}{a f(t)} \right)\right]\right) },$\\
   
      &&$k=\frac{-3a}{c}$&$B= \frac{-3a}{c}\frac{f(t)}{r} A$\\
      \hline
      $bG_2+cG_3$ &$A=h(t)r^{(b+c)/c},$ &$b= \left(\pm\frac{\sqrt{3}}{3} -1\right)c$&$A=\pm\frac{\sqrt{3} }{2  }\frac{g'\sqrt{g} }{\sqrt{g}+ d}r^{ \pm\frac{\sqrt{3}}{3} },$\\
   
      &$B=g(t)r^{b/c}, \text{ and }t$&&$B= gr^{ \pm\frac{\sqrt{3}}{3} -1 }$\\
     
     % \cline{3-4}

\hline

$aG_1+bG_2$ &$A=h(r)\frac{\exp\left(\int ^t \frac{bdt}{af(t)} \right) }{f(t)},$ &$g=kh$&$A=\frac{d}{r^{4/3}f(t)}\left( \frac{\left( \alpha (r)+2 a^2+a b k r\right)^{2} }{ \left(\alpha (r)+2a b k r+a^2\right) }\right)^{1/3} \exp \left(\int ^t \frac{bdt}{af(t)} -\frac{ \alpha (r)}{3 a^2}-\frac{b k r}{3 a}\right),$\\
          
    &$B=g(r)  $&&$B=kf(t) A$\\
           
           \cline{4-4}
            &$\times\exp\left(\int ^t \frac{bdt}{af(t)}\right),$& & $A=\frac{d}{f(t)}\left( \frac{\alpha (r)+2 a b k r+a^2 }{\left( \alpha (r)+2 a^2+a b k r\right)^{2}}\right)^{1/3} \exp \left( \int ^t \frac{bdt}{af(t)}+\frac{\alpha (r)}{3 a^2}-\frac{b k r}{3 a}\right),$\\
           
            &$\text{ and } r$&&$B=kf(t) A$\\
          
 \hline

\end{tabular*}
\end{table*}

\begin{table*}[T]
\centering
\caption{ Equation of state \label{table2} }
\begin{tabular*}{\textwidth}{@{\extracolsep{\fill}}lll@{}}
                                                                          
    \hline
Symmetry&       Gravitational potentials&Equation of state \\
    \hline
     
    $aG_1+bG_2$ &   $A=\frac{1}{f(t)} \left(m \left[ r^{-1/c}\exp\left( \int ^t \frac{dt}{a f(t)} \right)\right]^{\frac{a \sqrt{3 a+c k}}{2 k \sqrt{a-c k}}}\right.$&$p_ \parallel=\lambda\mu, $\\

    $\qquad+cG_3$&$\qquad \left. +n \left[ r^{-1/c}\exp\left( \int ^t\frac{dt}{a f(t)} \right)\right]^{-\frac{a \sqrt{3 a+c k}}{2 k \sqrt{a-c k}}}\right){}^{\frac{2 c k}{3 a+c k}},$&$\lambda =\frac{a c k}{a^2-a c k+c^2 k^2}$\\
   
&$B=\frac{kf(t)}{r} A$&\\

  \cline{2-3}
     &$A=\frac{n}{f(t)}  \frac{  \left[ r^{-1/c}\exp\left( \int^t \frac{dt}{a f(t)} \right)\right]^{m}}{\exp\left( \frac{c^2}{48}  \log ^2 \left[ r^{-1/c}\exp\left( \int ^t \frac{dt}{a f(t)} \right)\right]\right) },$&$p_ \parallel=\lambda\mu, $\\
   
      &$B= \frac{-3a}{c}\frac{f(t)}{r} A$&$\lambda =-\frac{3}{13}$\\
      \hline
      $bG_2+cG_3$ & $A=\pm\frac{\sqrt{3} }{2  }\frac{g'\sqrt{g} }{\sqrt{g}+ d}r^{ \pm\frac{\sqrt{3}}{3} },$&$p_ \parallel=\lambda\mu, $\\
   
      &$B= gr^{ \pm\frac{\sqrt{3}}{3} -1 }$&$\lambda =-\frac{2}{7}$\\
     
    \hline
 \end{tabular*}
\end{table*}

\section*{ACKNOWLEDGEMENTS}
\noindent  GZA and KSG thank the University of KwaZulu--Natal for continuing support.
SDM   acknowledges that this research is supported by the South African Research Chair Initiative of the Department of Science and Technology and the National Research Foundation.

\end{document}